\documentstyle[12pt]{article}

\font\mybb=msbm10 at 11pt
\def\bb#1{\hbox{\mybb#1}}
\def\ZZ{\bb{Z}}

\begin{document}
\thispagestyle{empty}
{\baselineskip=12pt
\hfill CALT-68-1994

\hfill hep-th/9505170

\hfill May 1995

\vspace{1.0cm}}
\centerline{\large \bf Classical Duality
Symmetries in Two Dimensions
\footnote{Work supported in part by the U.S. Dept. of Energy
under Grant No. DE-FG03-92-ER40701.}}
\bigskip
\bigskip
\centerline{\large John H. Schwarz\footnote{Email: jhs@theory.caltech.edu}}
\medskip
\centerline{\it California Institute of Technology, Pasadena, CA 91125, USA}
\bigskip
\bigskip
\bigskip
\centerline{{\it Presented at the Conference
``Strings `95: Future Perspectives in String Theory''}}

\centerline{{\it University of Southern California \quad March 1995}}
\vskip 1.0 truein
\parindent=1 cm

\begin{abstract}
Many two-dimensional classical field theories have
hidden symmetries that form an infinite-dimensional
algebra.  For those examples that correspond to effective
descriptions of compactified superstring theories, the duality group is
expected to be a large discrete subgroup of the hidden symmetry group.  With
this motivation, we explore the hidden symmetries of principal chiral models
and symmetric space models.
\end{abstract}
\vfil\eject
\setcounter{page}{1}
\section{Introduction}
Despite the impressive progress that has been achieved in understanding string
theory during the past decade, the theory has not yet been satisfactorily
formulated.  There are rules for identifying classical solutions, but we do not
know the equation that these ``solutions'' solve.  There are also rules for how
to compute quantum corrections to these classical solutions to any finite order
in perturbation theory, but we do not know how to compute non-perturbative
quantum effects.  Recently, a window onto non-perturbative string theory has
begun to open with the discovery of various duality symmetries and mappings
that
can relate weak coupling and strong coupling\cite{font}.
It seems possible that by
developing a deeper understanding of these dualities we will eventually be led
to the long-sought non-perturbative formulation of the theory.  More
specifically, there is ever increasing evidence that at a fundamental level
there is just one superstring theory\cite{hull}.
It seems reasonable that it should be
largely characterized by its group of symmetries.  This group should contain
all the duality symmetries, which are to be viewed as gauge symmetries in the
sense that they relate equivalent configurations that should be counted only
once in the path integral that defines the theory.

The specific duality groups that have been identified to date should be viewed
as subgroups of the complete duality group of string theory.  The point is
that,
with currently available techniques, it is only possible to study superstrings
in specific classical backgrounds, for example ones in which some spatial
dimensions are compactified on a particular manifold.  For any such choice,
some of the duality symmetries are easily identified and others are completely
hidden.  The ``visible'' ones are those that can be understood in terms of the
massless modes whereas the rest must involve massive modes in a complicated
way.  In particular, the visible group in question is realized nonlinearly on
the massless scalars.  The effective classical theory has a continuous
symmetry group,
but string and quantum effects restrict it to a discrete duality group.  Thus,
for example, the heterotic string toroidally compactified to four dimensions
has an $O(6,22;\ZZ)$ T-duality group and an $SL(2,\ZZ)$ S-duality group.  In
three dimensions these are combined and extended to give an $O(8,24;\ZZ)$
U-duality group\cite{senb}.
These examples illustrate an important point: the more
dimensions are compactified, the more duality symmetries become visible.  Since
my goal is to understand the complete group, this suggests that it should be
worthwhile to consider cases in lower dimensions in which even larger symmetry
groups can be identified.  In the case of toroidally compactified type II
superstrings in d dimensions, the duality groups appear to
be integral subgroups of maximally
noncompact forms of $E_{11-d}$.  If this rule extrapolates correctly, it would
give $E_9$ (the affine extension of $E_8$) in two dimensions and $E_{10}$ (a
hyperbolic Lie algebra) in one dimension\cite{nicolaia}.
It may be that affine Lie algebras
are generic in two dimensions and hyperbolic ones are generic in one dimension.
 As a modest first step to see if this is the case, I have investigated the
affine symmetry algebras of certain classes of two dimensional theories,
principal chiral models and symmetric space models.  The principal chiral
models are a warm-up exercise, whereas symmetric space models are relevant to
our problem.  However, to further simplify matters I have omitted gravity,
fermions, and quantum effects.  The results presented here summarize a recent
paper to which I refer the reader for additional details\cite{mypaper}.

In Ref. [5], I attempted to sketch the history of the study of hidden
symmetries in two-dimensional models.  Here I will simply remark that
relativists, beginning with Geroch\cite{geroch}
in 1971, studied hidden symmetries of
classical theories coupled to gravity in two dimensions, which corresponds to
four-dimensional Einstein theory with two commuting Killing vectors.  The
symmetries enable one to construct new solutions out of old ones.  Field
theorists, on the other hand, studied two-dimensional quantum theories (mostly
without gravity) as simpler analogs of four-dimensional gauge theories.  Hidden
non-local conserved charges of principal chiral models were discovered by
Pohlmeyer and L\"uscher\cite{pohlmeyer}.  The algebra of
the corresponding symmetry transformations
was studied by Dolan, Wu, and others\cite{dolana}.
Hidden symmetries in the supergravity
context have been explored most notably by Julia, Breitenlohner, Maison, and
Nicolai\cite{julia,breitenlohner,nicolaic}. Preliminary studies of these
symmetries for 2D string theory have been given by Bakas and
Maharana\cite{bakas}.
A more detailed discussion for
the toroidally compactified heterotic string has been given
by Sen\cite{sentwod}.

\section{Principal Chiral Models}

Principal chiral models (PCM's) are based on fields $g(x)$ that map space-time
into a group manifold $G$, which we may assume to be compact.  Even though
these models are not directly relevant to the string theory and supergravity
applications that we have in mind, they serve as a good warm-up exercise, as
well as being of some interest in their own right.  Symmetric space models,
which are relevant, share many of the same features, but are a little more
complicated.  They will be described in the next section.

The classical theory of PCM's, in any dimension, is defined by the lagrangian
\begin{equation}
{\cal L} = \eta^{\mu\nu} tr (A_\mu A_\nu),
\end{equation}
where the connection $A_\mu$ is defined in terms of the group variables by
\begin{equation}
A_\mu = g^{-1} \partial_\mu g = \sum A_\mu^i T_i.
\end{equation}
Here $\eta^{\mu\nu}$ denotes the Minkowski metric for flat space-time, and the
$T_i$ are the generators of the Lie algebra,
\begin{equation}
[T_i,T_j] = f_{ij}{}^k T_k.
\end{equation}
They may be taken to be matrices in any convenient representation.  The
classical equation of motion is derived by letting $\delta g$ be an arbitrary
infinitesimal variation of $g$ for which $\eta = g^{-1}\delta g$
belongs to the Lie algebra ${\cal G}$.  Under this variation
\begin{equation}
\delta A_\mu = D_\mu \eta = \partial_\mu \eta + [A_\mu, \eta],
\end{equation}
and thus
\begin{equation}
\delta {\cal L} = 2\, {\rm tr} (A^\mu D_\mu \eta) = 2\, {\rm tr}
(A^\mu \partial_\mu \eta).
\end{equation}
{}From this it follows that the classical equation of motion is
\begin{equation}
\partial_\mu A^\mu = 0,
\end{equation}
as is well-known.  Since $A_{\mu}$ is pure gauge, the Bianchi identity is
\begin{equation}
F_{\mu\nu} = \partial_\mu A_\nu - \partial_\nu A_\mu + [A_\mu, A_\nu] = 0.
\end{equation}

The PCM in any dimension has manifest global $G\times G$ symmetry corresponding
to left and right group multiplication.  Remarkably, in two dimensions this is
just a small subgroup of a much larger group of ``hidden'' symmetries.  To
describe how they arise, it is convenient to introduce light-cone coordinates
\begin{equation}
x^\pm = x^0 \pm x^1,\quad \partial_\pm =
{1\over 2} (\partial_0 \pm \partial_1).
\end{equation}
Expressed in terms of these coordinates, the equation of motion and Bianchi
identity take the forms
\begin{equation}
\partial_{\mu} A^{\mu} = \partial_+ A_- + \partial_- A_+ = 0
\end{equation}
\begin{equation}
F_{+-} = \partial_+ A_- - \partial_- A_+ + [A_+, A_-] = 0.
\end{equation}

A standard technique (sometimes called the ``inverse scattering method'') for
discovering the ``hidden symmetries'' of integrable models, such as a PCM in
two dimensions, begins by considering a pair of linear differential equations,
known as a Lax pair.  In the present context the appropriate equations are
\begin{equation}
(\partial_+ + \alpha_+ A_+) X = 0 \quad {\rm and} \quad
(\partial_- + \alpha_- A_-) X = 0,
\end{equation}
where $\alpha_\pm$ are constants.
These equations are compatible, as a consequence
of equations (9) and (10), provided that
\begin{equation}
\alpha_+ + \alpha_- = 2 \alpha_+ \alpha_-.
\end{equation}
It is convenient to write the solutions to this equation in terms of a
``spectral parameter'' $t$ in the form
\begin{equation}
\alpha_+ = {t\over t - 1} , \quad \alpha_- = {t\over t + 1}.
\end{equation}

The variable $X$ in eqs. (11) is a group-valued function of the space-time
coordinate, as well as the spectral parameter.  The integration constant can be
fixed by requiring that $X$ reduces to the identity element of the group at a
``base point'' $x_0^\mu$.  A formal solution to eqs. (11) is then given by a
path-ordered exponential
\begin{equation}
X (x, t) = P \exp \Big\{ - \int_{x_{0}}^x (\alpha_+ A_+ dy^+
+ \alpha_- A_- dy^-)\Big\},
\end{equation}
where the path ordering has $x$ on the left and $x_0$ on the right.  The
integral is independent of the contour provided the space-time is simply
connected.  This is the case, since we are assuming a flat Minkowski
space-time.  If one were to choose a circular spatial dimension instead, the
multivaluedness of $X$ would raise new issues, which we will not consider here.
Note that $X$ is group-valued for any real $t$, except for
the singular points $t = \pm 1$.

The next step is to consider the variation $g^{-1} \delta g = \eta$, with
\begin{equation}
\eta (\epsilon, t) = X(t) \epsilon X(t)^{-1},
\end{equation}
where $\epsilon = \sum \epsilon^i T_i$ and $\epsilon^i$ are infinitesimal
constants.  The claim is that the variation $\delta(\epsilon, t) g = g\eta$
preserves the equation of motion $\partial \cdot A = 0$ and, therefore,
describes symmetries of the classical theory.  To show this, one simply notes
that the Lax  pair implies that
\begin{equation}
\delta A_\pm = D_\pm \eta = \partial_\pm \eta + [A_\pm, \eta] = \pm {1\over t}
\partial_\pm \eta,
\end{equation}
and, therefore, $\partial \cdot (\delta A) = 0$ as required.

Let us now consider the commutator of two symmetry transformations $[\delta
(\epsilon_1, t_1), \delta (\epsilon_2, t_2)]$.  The key identity that is
required is
\begin{equation}
\delta (\epsilon_1, t_1) X (t_2) = {t_2\over t_1 - t_2} (\eta (\epsilon_1, t_1)
X (t_2) - X(t_2) \epsilon_1).
\end{equation}
Identities such as this are used frequently in this work.  The method of proof
is always the same.  One shows that both sides of the equation satisfy the same
pair of linear differential equations and boundary conditions and then
concludes by uniqueness that they must be equal.  The required equations are
obtained by varying the Lax pair.  Using this identity it is easy to derive
\begin{equation}
[\delta (\epsilon_1, t_1), \delta (\epsilon_2, t_2)] = {t_1 \delta
(\epsilon_{12}, t_1) - t_2 \delta (\epsilon_{12}, t_2)\over t_1 - t_2},
\end{equation}
where
\begin{equation}
\epsilon_{12} = [\epsilon_1, \epsilon_2] = f_{ij}{}^k
\epsilon_1^i \epsilon_2^j T_k.
\end{equation}

In order to understand the relationship between the algebra (18) and current
algebra associated with the group $G$, we need to do some sort of mode
expansion with respect to the parameter $t$.  The standard approach in the
literature is to do a power series expansion in $t$, $\delta (\epsilon, t) =
\sum_{n = 0}^\infty \delta_n (\epsilon) t^n$, identifying the $\delta_n
(\epsilon)$ as distinct symmetry transformations.  This gives half of a current
algebra:
\begin{equation}
[\delta_m (\epsilon_1), \delta_n (\epsilon_2)] = \delta_{m + n} (\epsilon_{12})
\quad m,n \geq 0 .
\end{equation}
Actually, $\delta (\epsilon, t)$ contains more information than is extracted in
this way, and in Ref. [5] I found a nice way to reveal it.  The idea is to
define variations $\Delta_n (\epsilon) g$ for all integers $n$ by the contour
integral
\begin{equation}
\Delta_n (\epsilon) g = \int_{\cal C} {dt\over 2\pi i} t^{-n -1}
\delta (\epsilon, t) g  \quad  n \in \ZZ,
\end{equation}
where the contour ${\cal C} = {\cal C}_+ + {\cal C}_-$ and ${\cal C}_\pm$ are
small clockwise circles about $t = \pm 1$.  By distorting contours it is easy
to show that $\Delta_n (\epsilon) = \delta_n (\epsilon)$ for $n > 0$.  The
negative integers $n$ are given entirely by  poles at $t = \infty$.  In other
words, they correspond to the coefficients in a series expansion in inverse
powers of $t$.  $\Delta_0$ receives contributions from
poles at both $t = 0$ and $t =
\infty$.  (Explicitly, $\Delta_0 (\epsilon) g = [g, \epsilon]$.)  Because
$g^{-1} \Delta_n g$ can be related to such series expansions, it is clear that
it is Lie-algebra valued.\footnote{If one tried to define further symmetries
corresponding to the contours ${\cal C}_\pm$ separately or by allowing $n$ to
be non-integer, the transformations defined in this way would also appear to
preserve $\partial \cdot A = 0$.  However, these could fail to be honest
symmetries because $g^{-1} \delta g$ might not be Lie-algebra valued.}

Using the definition (21) and the commutator (18), it is an easy application of
Cauchy's theorem to  deduce the affine current algebra (without center)
\begin{equation}
[\Delta_m (\epsilon_1), \Delta_n (\epsilon_2)] = \Delta_{m+n} (\epsilon_{12})
\qquad m,n \in {\ZZ}.
\end{equation}
Equivalently, in terms of charges we have
\begin{equation}
[J_m^i, J_n^j] = f^{ij}{}_k J_{m+n}^k.
\end{equation}

Having found current algebra symmetries for classical PCM's, it is plausible
that they should also have Virasoro symmetries\cite{cheng}.
We now show that,  modulo an
interesting detail, this is indeed the case.  Since the infinitesimal parameter
in this case is not Lie-algebra valued, it can be omitted without ambiguity.
With this understanding, the Virasoro-like transformation
\begin{equation}
\delta^V (t) g =  g ((t^2 - 1) \dot X(t) X(t)^{-1} + I),
\end{equation}
where the dot denotes a $t$ derivative and
\begin{equation}
I = \dot X(0) = \int_{x_{0}}^x (A_+ dy^+ - A_- dy^-),
\end{equation}
can be shown to be an invariance of the equation of motion $\partial \cdot A =
0$.  We can extract modes $\delta_n^V$, for all integers $n$, by the same
contour integral definition used above
\begin{equation}
\delta_n^V g = \int_{\cal C} {dt\over 2\pi i} t^{-n -1} \delta^V (t) g.
\end{equation}
Again, contour deformations give pole contributions at $t = 0$ and $t = \infty$
only, and therefore, one sees that $g^{-1} \delta_n^V g$ is Lie-algebra valued.

The analysis of the algebra proceeds in the same way as for the current
algebra,
though the formulas are quite a bit more complicated.  For example, commuting
a Virasoro symmetry transformation with a current algebra symmetry
transformation gives
\[[\delta^V (t_1), \delta (\epsilon, t_2)] g =  \Big({1\over t_2}
(\delta (\epsilon, 0) - \delta (\epsilon, t_2))\]
\begin{equation}
\quad + {t_2 (t_1^2 - 1)\over (t_1 - t_2)^2} (\delta (\epsilon, t_1)
- \delta (\epsilon, t_2))
+ {t_1 (1 - t_2^2)\over t_1 - t_2} ~ {\partial\over\partial t_2} \delta
(\epsilon, t_2) \Big) g .
\end{equation}
Using eq. (27) and the contour integral definitions, one finds after some
integrations by parts and use of Cauchy's theorem that
\begin{equation}
[\delta_{m}^V, \Delta_{n} (\epsilon)] g = n \int_{\cal C} {dt\over
2\pi i} t^{-m -n -2} (t^2 -1) \delta (\epsilon, t) g.
\end{equation}

Now let us re-express the algebra in terms of charges $J_n^i$ (as before) and
$K_m$ (corresponding to $\delta_m^V$). In this notation, eq. (28) becomes
\begin{equation}
[K_{m}, J^i_{n}] = n (J^i_{m + n-1} - J^i_{m +
n + 1}).
\end{equation}
This formula is to be contrasted with what one would expect for the usual
Virasoro generators $L_n$
\begin{equation}
[L_{m}, J^i_{n}] = - n J^i_{m + n}.
\end{equation}
Comparing equations, we see that we can make contact with
the usual (centerless)
Virasoro algebra if we identify
\begin{equation}
K_n = L_{n + 1} - L_{n - 1}.
\end{equation}
However, it should be stressed that we have only defined the differences $K_n$
and not the individual $L_n$'s.  Still, this identification is useful since it
tells us that
\begin{equation}
[K_m, K_n] = (m - n) (K_{m+n+1} - K_{m+n-1}).
\end{equation}

Let us see what happens if we try to construct the $L_n$'s.  The easiest
approach is to define $K(\sigma) = \sum_{-\infty}^\infty K_n e^{in\sigma}$ and
$T(\sigma) = \sum_{-\infty}^\infty L_n e^{in\sigma}.$  Then eq. (31) implies
that
\begin{equation}
T(\sigma) = {i\over 2} ~ {K(\sigma)\over\sin \sigma}.
\end{equation}
The remarkable fact is that $K(\sigma)$ does not vanish at $\sigma = 0$ and
$\sigma = \pi$.  Therefore, $T (\sigma)$ diverges at these points and the
individual $L_n$'s do not exist.  The integrals that would define them are
logarithmically divergent.

\section{Symmetric Space Models}

An interesting class of integrable two-dimensional models consists of theories
whose fields map the space-time into a symmetric space.  Let $G$ be a simple
group and $H$ a subgroup of $G$.  Then the Lie algebra ${\cal G}$ can be
decomposed into the Lie algebra ${\cal H}$ and its orthogonal complement ${\cal
K}$, which contains the generators of the coset $G/H$.  The coset space $G/H$
is called a symmetric space if $[{\cal K, K}]\subset {\cal H}$, in other words
the commutators of elements of ${\cal K}$ belong to ${\cal H}$.  The examples
that arise in string theory and supergravity are non-compact symmetric space
models (SSM's), such as those mentioned in the introduction.  For such models,
$G$ is a non-compact Lie group and $H$ is its maximal compact subgroup.  The
generators of ${\cal H}$ are antihermitian and those of ${\cal K}$ are
hermitian.  Therefore, since the commutator of two hermitian matrices is
antihermitian, $[{\cal K, K}]\subset {\cal H}$ and $G/H$ is a (non-compact)
symmetric space.

Symmetric space models can be formulated starting with arbitrary $G$-valued
fields, $g(x)$, like those of PCM's.  To construct an SSM, we associate local
$H$ symmetry with left multiplication and global $G$ symmetry with right
multiplication.  Thus, we require invariance under infinitesimal
transformations of the form
\begin{equation}
\delta g = - h (x) g +
g \epsilon \quad h\in {\cal H}, \  \epsilon\in {\cal G}.
\end{equation}
The local symmetry effectively removes $H$ degrees of freedom so that only
those of the coset remain.  The next step is to define
\begin{equation}
P_\mu = {1\over 2} (g \partial_\mu g^{-1} + \partial_\mu g^{-1\dagger}
g^{\dagger}),
\end{equation}
\begin{equation}
A_\mu = - 2 g^{-1} P_\mu g.
\end{equation}
and to observe that this $A_\mu$ is invariant under local $H$ transformations.
It can be recast in an alternative form that makes this manifest, specifically
\begin{equation}
A_\mu = M^{-1} \partial_\mu M,
\end{equation}
where
\begin{equation}
M = g^\dagger g.
\end{equation}
Note that $g$ and $M$ are analogous to a vielbein and metric in general
relativity.  $M$ parametrizes the symmetric space $G/H$ without extra degrees
of freedom.  In the case of a compact SSM the factor
$g^\dagger$ in the definition of
$M$ must be generalized to a quantity $\tilde{g}$, which is described in Ref.
[5].  Since $A_\mu$ is pure gauge, its field strength vanishes
\begin{equation}
F_{\mu\nu} = \partial_\mu A_\nu - \partial_\nu A_\mu + [A_\mu, A_\nu] = 0.
\end{equation}
The lagrangian is ${\cal L} = {\rm tr} (A^\mu A_\mu)$
and the classical equation of motion is
\begin{equation}
\partial^{\mu} A_{\mu} = 0.
\end{equation}
These formulas look the same as for PCM's, but $A_\mu$ is given in terms of
$g(x)$ by a completely different formula
(eqs. (37) and (38) instead of eq. (2).

In two dimensions we once again have the Bianchi identity  $F_{+-} = 0$ and the
equation of motion $\partial_+ A_- + \partial_- A_+ = 0$.  Therefore, it is
natural to investigate whether the formulas that gave rise to symmetries of
PCM's also gives rise to symmetries in this case.  With this motivation, we
once again form the Lax pair of equations
\begin{equation}
(\partial_\pm + \alpha_\pm A_\pm) X = 0,
\end{equation}
and note that they are compatible if we write $\alpha_\pm$ in terms of a
spectral parameter as in eq. (13).  Then the solution is given by the contour
independent integral
\begin{equation}
X(t) = P \exp \Big(- \int_{x_{0}}^x (\alpha_+ A_+ dy^+
+\alpha_- A_- dy^-) \Big),
\end{equation}
as before.  The obvious guess is that, just as for PCM's, the hidden
symmetry is described by
\begin{equation}
\delta g = g X(t) \epsilon X(t)^{-1} .
\end{equation}
This turns out to be correct.  Under an arbitrary infinitesimal variation
$g^{-1} \delta g = \eta (x) \in {\cal G}$, we have
\begin{equation}
\delta M = \eta^\dagger M + M \eta ,
\end{equation}
which implies that
\begin{equation}
\delta A_\mu = D_\mu \eta + D_\mu (M^{-1} \eta^\dagger M).
\end{equation}
The first term is the same as for a PCM, but the second one is new.  The
symmetry requires that $\partial^\mu (\delta A_\mu) = 0$, when we substitute
$\eta = X \epsilon X^{-1}$.  The vanishing of the contribution from the first
term in eq. (45) is identical to the PCM case.  The second term in eq. (45)
also has a vanishing divergence (for $\eta = X \epsilon X^{-1}$).

Next, we wish to study the algebra of these symmetry transformations.  The
commutator turns out to be
\begin{equation}
[\delta(\epsilon_1, t_1), \delta(\epsilon_2, t_2)]g =
{t_1 \delta (\epsilon_{12}, t_1) -
t_2 \delta (\epsilon_{12}, t_2)\over t_1 - t_2}g +\delta^{\prime}g
+ \delta^{\prime\prime}g ,
\end{equation}
where the first term is the same as we found for PCM's, but there are two
additional pieces.  The $\delta' g$ term is a local ${\cal H}$ transformation
of the form $h (x) g$, which is a symmetry of the theory.  It is trivial in its
action on $M = g^\dagger g$, which is all that appears in ${\cal L}$.  (This is
analogous to the trivial invariance of the Einstein--Hilbert action under local
Lorentz transformations.)  The $\delta^{\prime\prime} g$
term is given by
\begin{equation}
\delta^{\prime\prime}g = {t_1 t_2\over 1 - t_1 t_2 }
\left( \delta(\epsilon'_{12}, t_2) - \delta(\epsilon'_{21},t_1)\right),
\end{equation}
where
\begin{equation}
\epsilon'_{12} = M_0^{-1} \epsilon_1^{\dagger} M_0 \epsilon_2
-\epsilon_2 M_0^{-1}\epsilon_1^{\dagger} M_0,
\end{equation}
and $M_0 = M(x_0)$.

As in the PCM, we define modes by contour integrals of the form given in eq.
(21), and associate charges $J_n^i$ to the transformation $\Delta_n
(\epsilon)$.  These can be converted to ``currents'' $J_n^i (\sigma) = \sum
e^{in\sigma} J_n^i$.  In the case of an SSM, there are two distinct classes of
currents, those belonging to ${\cal H}$ and those belonging to ${\cal K}$.  As
Ref. [5] shows in detail, the significance of the $\delta^{\prime\prime} g$
term in eq. (46) is that the ${\cal H}$ currents satisfy Neumann boundary
conditions at the ends of the interval $0 \leq \sigma \leq \pi$, while the
${\cal K}$ currents satisfy Dirichlet boundary conditions at the two ends
\renewcommand{\theequation}{49a}
\begin{equation}
J^{i\prime} (0) = J^{i\prime} (\pi) =0 \quad{\rm for}
\quad J^i \in {\cal H}
\end{equation}
\renewcommand{\theequation}{49b}
\begin{equation}
J^i (0) =J^i (\pi) = 0 \quad{\rm for}\quad J^i \in {\cal K}.
\end{equation}
As a result, $J_n^i = J_{-n}^i$ for ${\cal H}$ charges and
$J_n^i = - J_{-n}^i$ for
${\cal K}$ charges.  Thus the mode expansions become
\renewcommand{\theequation}{50a}
\begin{equation}
J^i (\sigma) = J_0^i + 2 \sum_{n=1}^\infty \cos n \sigma
J_n^i \quad{\rm for}\quad J^i \in {\cal H}
\end{equation}
\renewcommand{\theequation}{50b}
\begin{equation}
J^i (\sigma) = 2 i\sum_{n=1}^\infty \sin n \sigma J_n^i \quad{\rm for}
\quad J^i \in {\cal K}.
\end{equation}
In terms of the modes, local current algebra on the line segment $0 \leq \sigma
\leq \pi$ then implies that
\renewcommand{\theequation}{51a}
\begin{equation}
[J_m^i, J_n^j] = f^{ij}{}_k (J_{m + n}^k + J_{m-n}^k) \quad{\rm for}\quad
J_n^j \in {\cal H}
\end{equation}
\renewcommand{\theequation}{51b}
\begin{equation}
[J_m^i, J_n^j] = f^{ij}{}_k (J_{m+n}^k - J_{m-n}^k) \quad{\rm for}
\quad J_n^j \in {\cal K}.
\end{equation}
\renewcommand{\theequation}{\arabic{equation}}
I propose to call this kind of a current algebra $\hat G_H$.
\setcounter{equation}{51}

This result is somewhat surprising, because it seems to conflict with claims in
the literature that the symmetry is an ordinary $\hat G$ current algebra on a
circle, like what we found for PCM's.  Actually, a few authors did
previously obtain the same $\hat G_H$ algebra for a special  class of SSM's,
though they did not offer an interpretation\cite{wu}.
As an additional check on the
result, Ref. [5] studies the formulation of PCM's as SSM's based on the coset
$G\times G/G$ \cite{chau}.
It shows that the $\hat G_H$ symmetry of this model
agrees with $\hat G$, the symmetry of the PCM.  If the symmetry of a G/H SSM
were a full $\hat G$ (rather than the subgroup $\hat G_H$),
then the SSM construction of a PCM would imply that it has a $\hat G
\times \hat G$ symmetry.  Such symmetries occur for WZNW models, of course, but
there are no Wess--Zumino terms in our models.

The Virasoro-like symmetries of PCM's also generalize to SSM's.  The natural
guess is that, just as for the current algebra symmetry, the same formula will
describe the symmetry in this case, namely
\begin{equation}
\delta^V (t) g =
g \Big((t^2 - 1) \dot X (t) X (t)^{-1} + I\Big).
\end{equation}
This turns out to be correct, but once again the algebra differs from that of
PCM's.  We find that
\begin{equation}
[\delta^V (t_1), \delta(\epsilon, t_2)] g =
\delta g + \delta' g + \delta^{\prime\prime} g,
\end{equation}
where $\delta g$ is the PCM result given in eq. (27).  The $\delta' g$ is a
local ${\cal H}$ transformation and $\delta^{\prime\prime} g$ contains new
terms.  (The formulas are given in Ref. [5].)  The crucial question becomes
what $\delta^{\prime\prime} g$ contributes $[\delta_m^V, \delta_n
(\epsilon)]g$, when we insert it into the appropriate contour integrals, or,
equivalently, what it contributes to $[K_m, J_n^i]$.  The result is
\begin{equation}
[K_m, J_n^i] = n (J_{m+n-1}^i - J_{m+n+1}^i - J_{n-m+1}^i + J_{n-m-1}^i).
\end{equation}
The first two terms are the PCM result of eq. (29), while the last two terms
are the new contribution arising from $\delta^{\prime\prime} g$.

After our experience with the current algebra symmetry, the interpretation of
the result (54) is evident.  The generators $K_m$ satisfy the restrictions $K_m
= K_{-m}$, just like the ${\cal H}$ currents.  In other words, $K(\sigma)$
satisfies Neumann boundary conditions at the ends of the interval $0 \leq
\sigma \leq \pi$.  Just as for PCM's, one can define a stress tensor
\begin{equation}
T(\sigma) = {i\over 2} {K(\sigma)\over \sin \sigma},
\end{equation}
which satisfies the standard stress tensor algebra.
As before, it is singular at
$\sigma=0$ and $\sigma=\pi$, so that modes $L_m$ do not exist.

\section{Concluding Remarks}

We have seen that the key ingredient in the study of hidden symmetries of
two-dimensional integrable models
is the group element obtained by integrating the Lax pair
\begin{equation}
X (x, t) = P \exp \Big\{ - \int_{x_{0}}^x (\alpha_+ A_+ dy^+
+ \alpha_- A_- dy^-)\Big\}.
\end{equation}
Since it plays such a central role, it is natural to explore what happens if
one makes a change of variables
\begin{equation}
g'(x) = g(x) X (x, u) , \quad - 1 < u < 1.
\end{equation}
The result is quite different for PCM's and SSM's.  In the case of PCM's, it
turns out the $g'$ equation of motion is
\begin{equation}
(1 - u) \partial_+ A'_- + (1 + u) \partial_- A'_+ = 0.
\end{equation}
This is recognized to be the equation of motion obtained from the action
\begin{equation}
S_u (g') = S_{PCM} (g') + u S_{WZ} (g'),
\end{equation}
where $S_{WZ}$ denotes a Wess--Zumino term.  Thus we learn that all values of
$u$ other than $u = \pm 1$ give equivalent classical theories.  The special
values $u = \pm 1$, which are different, correspond to WZNW theory.  For the
quantum theory the normalization matters, and one should consider
\begin{equation}
S_{k,u}= k\Big( {1 \over u} S_{PCM} + S_{WZ}\Big),
\end{equation}
where $k$ is an integer.  It is plausible that for a given $k$, all values of
$u$ other than $\pm 1$ give equivalent quantum theories.

In the case of SSM's the change of variables in eq. (57) does not give rise to
a WZ term.  Instead it is a (finite) symmetry transformation that corresponds
to exponentiating the infinitesimal symmetry generated by $K_0$.

There is much work that still remains to be done if the analysis presented here
is going to be applied to the study of string theory duality symmetries.
Obvious future directions include coupling the models to 2D gravity as well as
adding fermions and supersymmetrizing.  Other issues involve understanding how
quantization breaks the symmetry to discrete subgroups.  This requires dealing
with finite symmetry elements, rather than the infinitesimal elements described
here.  Considerable progress in addressing these issues has been made by Sen in
his study of the toroidally compactified heterotic string in two
dimensions\cite{sentwod}.
He identified discrete current algebra symmetries.  It would be interesting to
determine whether there are also discrete Virasoro symmetries.

Since compactification to one dimension is expected to give even larger
hyperbolic Lie algebra symmetry groups, that should be very interesting to
explore.  A first step, it seems to me, would be to understand how the
two-dimensional analysis is modified when the spatial dimension is a circle.
As we have pointed out already, the formula for $X$ in eq. (14) is no longer
single-valued in that case, so new issues arise.

I would like to acknowledge helpful discussions with Ashoke Sen.

\end{document}